%% file: sigconf-camera-ready.tex
\documentclass[sigconf]{acmart}
\input{_package.tex}
\input{_definition.tex}

\AtBeginDocument{%
  }

\copyrightyear{2025}
\acmYear{2025}
\setcopyright{cc}
\setcctype{by}
\acmConference[CIKM '25]{Proceedings of the 34th ACM International Conference on Information and Knowledge Management}{November 10--14, 2025}{Seoul, Republic of Korea}
\acmBooktitle{Proceedings of the 34th ACM International Conference on Information and Knowledge Management (CIKM '25), November 10--14, 2025, Seoul, Republic of Korea}\acmDOI{10.1145/3746252.3761164}
\acmISBN{979-8-4007-2040-6/2025/11}

\begin{document}


\title[Trustworthy AI Psychotherapy: Multi-Agent LLM Workflow for Counseling and Explainable \\ Mental Disorder Diagnosis]{Trustworthy AI Psychotherapy: Multi-Agent LLM Workflow for Counseling and Explainable Mental Disorder Diagnosis}

\author{Mithat Can Ozgun}\authornote{Equal contribution}
\orcid{0009-0006-2288-5796}
\affiliation{%
  \institution{Vrije Universiteit Amsterdam}
  \city{Amsterdam}
  \country{Netherlands}}
\email{m.c.ozgun@student.vu.nl}

\author{Jiahuan Pei}\authornotemark[1]\authornotemark[2]
\orcid{0000-0001-6951-8340}
\affiliation{%
  \institution{Vrije Universiteit Amsterdam}
  \city{Amsterdam}
  \country{Netherlands}}
\email{j.pei2@vu.nl} 

\author{Koen Hindriks}
\orcid{0000-0002-5707-5236}
\affiliation{%
  \institution{Vrije Universiteit Amsterdam}
  \city{Amsterdam}
  \country{Netherlands}}
\email{k.v.hindriks@vu.nl} 

\author{Lucia Donatelli}
\orcid{0000-0002-5974-7454}
\affiliation{%
  \institution{Vrije Universiteit Amsterdam}
  \city{Amsterdam}
  \country{Netherlands}}
\email{l.e.donatelli@vu.nl} 

\author{Qingzhi Liu}
\orcid{0000-0003-2621-9222}
\affiliation{%
  \institution{Wageningen University and Research}
  \city{Wageningen}
  \country{Netherlands}}
\email{qingzhi.liu@wur.nl} 


\author{Junxiao Wang}\authornote{Corresponding authors}
\orcid{0000-0001-7263-174X}
\affiliation{%
  \institution{Guangzhou University}
  \city{Guangzhou}
  \country{China}}
\email{junxiao.wang@gzhu.edu.cn} 

\renewcommand{\shortauthors}{Mithat Can Ozgun et al.}


\begin{abstract}
LLM-based agents have emerged as transformative tools capable of executing complex tasks through iterative planning and action, achieving significant advancements in understanding and addressing user needs. Yet, their effectiveness remains limited in specialized domains such as mental health diagnosis, where they underperform compared to general applications. Current approaches to integrating diagnostic capabilities into LLMs rely on scarce, highly sensitive mental health datasets, which are challenging to acquire. These methods also fail to emulate clinicians' proactive inquiry skills, lack multi-turn conversational comprehension, and struggle to align outputs with expert clinical reasoning. To address these gaps, we propose \textbf{\OurModel{}}, the first LLM-based agent workflow designed to autonomously generate DSM-5 Level-1 diagnostic questionnaires. By simulating therapist-client dialogues with specific client profiles, the framework delivers transparent, step-by-step disorder predictions, producing explainable and trustworthy results. This workflow serves as a complementary tool for mental health diagnosis, ensuring adherence to ethical and legal standards. Through comprehensive experiments, we evaluate leading LLMs across three critical dimensions: conversational realism, diagnostic accuracy, and explainability. Our datasets and implementations are fully open-sourced.
~\footnote{\url{https://github.com/mithatco/mental_health_multiagent}}

\end{abstract}




\maketitle

\section{Introduction}
\begin{figure}[htb!]
    \centering
    \includegraphics[width=\linewidth]{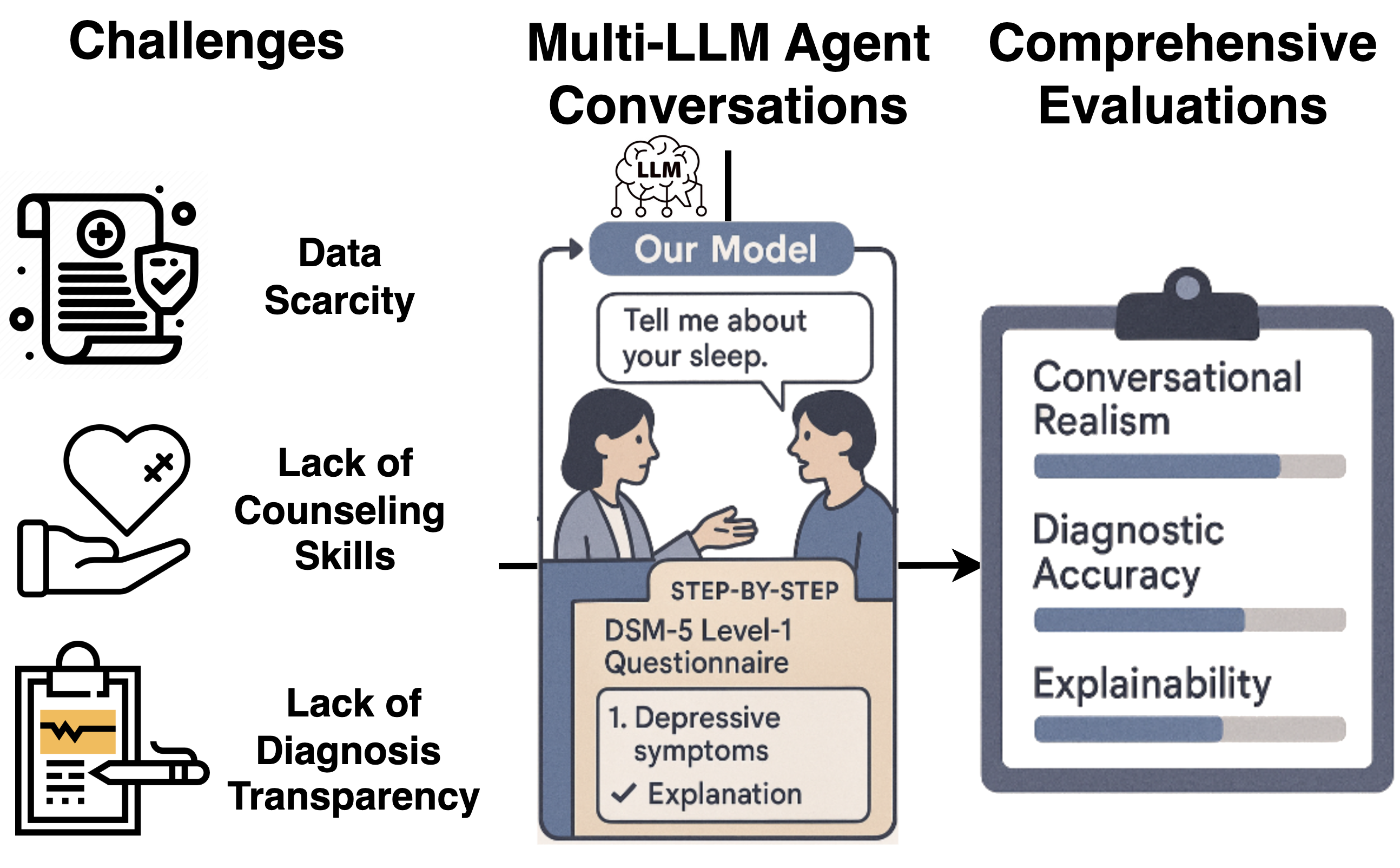}
    \caption{Illustration of the proposed workflow: trustworthy AI psychotherapy using multiple LLM agents for explainable mental disorder diagnosis.}
    \label{fig:motivation}
\end{figure}

\Ac{LLM}-based agents have rapidly advanced the automation of complex tasks, demonstrating strong capabilities in planning, reasoning, and adapting to user needs across diverse domains ~\cite{wu2025survey,na2025survey}. 
However, their application to specialized fields such as mental health diagnosis remains limited. 
In particular, LLMs often underperform in clinical contexts due to the scarcity of high-quality, sensitive mental health datasets and the unique requirements of diagnostic reasoning.

Existing approaches to LLM-based mental health assessment primarily focus on automating diagnostic interviews and generating synthetic clinical data to address privacy and data scarcity concerns~\cite{DeDuro2025, Yin2024, DiaSynthACL2025, SyntheticDataDepressionArXiv2024}. 
While these methods have enabled reproducible research and initial progress in clinical NLP, they often lack transparency, offer limited reasoning depth, and do not fully align with established clinical standards. 
Recent advances in multi-agent LLM workflows show promise in overcoming these barriers~\cite{ChenLiu2025, MultiAgentCognitiveBiasJMIR2024, SelfEvolvingMultiAgentArXiv2025}, but several critical technical challenges remain:
First, reliance on limited and private clinical data restricts scalability and reproducibility~\cite{Guo2024}. 
LLMs face challenges such as hallucination, difficulty with multi-turn dialogue, and limited ability to capture nuanced mental states, all of which hinder their effectiveness in clinical contexts~\cite{chung2023challenges, hua2024large}. 
Second, current methods often fail to emulate the proactive, multi-turn counseling strategies of clinicians, resulting in shallow or incomplete diagnostic reasoning~\cite{levkovich2025evaluating}.
Third, LLM outputs often diverge from expert clinical reasoning, producing diagnoses that are difficult to interpret or trust~\cite{Kafka2024}. 
The proprietary and opaque nature of LLM training further limits transparency, complicating their integration into clinical workflows.

In summary, despite the great potential of these early studies, the way they present workflow-related knowledge is often fragmented and inconsistent, and there is a lack of systematic research or rigorous benchmarks to evaluate the role of workflows. Therefore, there are still key challenges in formalizing, applying, and evaluating such knowledge for LLM-based agents in various real-worlds.

In real-world clinical settings, mental health triage often begins with brief screening tools such as the \emph{DSM‑5 Level‑1 Cross‑Cutting Symptom Measure} (a.k.a. the DSM-5 questionnaire) ~\cite{APA_DSM5_Level1}. 
These questionnaires are valued for their speed and practicality, helping clinicians quickly flag concerns and decide on next steps. Yet, this efficiency comes at a cost: clients often find the process opaque, unsure how their answers lead to diagnoses. 
This lack of transparency can erode trust and reduce engagement~\cite{Xu2024, Kim2025, Stade2024}. 
Studies show that when clients do not understand the reasoning behind assessments, they are less likely to share sensitive information or follow recommendations~\cite{Kafka2024,Thompson2007, Cirasola2024}. 
Clinicians, too, struggle to justify decisions based only on summary scores, especially in team-based care. 
The field urgently needs tools that combine the speed of questionnaires with clear, explainable logic for all users.

To address the above technical and practical limitations, we introduce \textbf{\OurModel{}}, a LLM-based multi-agent workflow that autonomously generates and administers DSM-5 Level-1 diagnostic questionnaires. 
Specifically, this workflow obtains one agent, primed with up-to-date clinical guidelines, and conducts the initial interview; 
a second agent, conditioned on a synthetic yet epidemiologically grounded disorder profile, plays the client role; 
and a third, diagnostician agent retrieves the relevant DSM-5 passages before formulating both a provisional diagnosis and a step-by-step explanation that cites individual utterances as evidence. 
It simulates realistic therapist–client dialogues using configurable client profiles, enabling transparent, step-by-step disorder prediction and explainable diagnostic rationales. 
This approach not only enhances the interpretability and trustworthiness of LLM-driven assessments but also ensures compliance with ethical and legal standards in mental health care. 
Because each stage is grounded in the same canonical source material, the resulting dialogue is not only coherent and empathetic but also internally self-documenting: 
Every answer flows into an explicit link with the criterion it supports—or contradicts.

The contributions of this work are threefold:
\begin{enumerate*}
    \item We present \OurModel{}, a novel multi-agent LLM workflow that generates and administers DSM-5 Level-1 diagnostic questionnaires, enhancing the interpretability and trustworthiness of LLM-driven assessments.
    \item We introduce a system which allows researchers to simulate realistic therapist–client dialogues, enabling transparent, step-by-step disorder prediction and explainable diagnostic rationales using custom profiles and questionnaires.
    \item We demonstrate the effectiveness of our approach through extensive experiments, showing its ability to produce coherent, empathetic dialogues while ensuring compliance with ethical standards in mental healthcare.
\end{enumerate*}


\section{Multi-Agent Interaction Workflow}\label{sec:method}

\begin{figure}[htb!]
  \centering
  \includegraphics[width=0.95\linewidth]{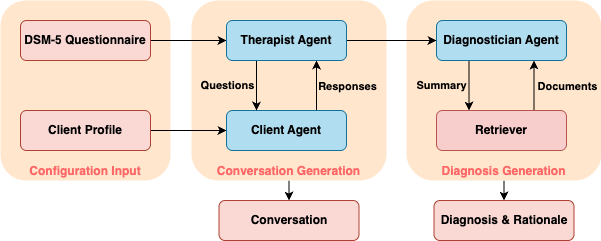}
  \caption{Multi-agent workflow. The Therapist Agent interviews the Client Agent.
   The resulting conversation transcript is passed to the diagnostician agent’, which produces a diagnosis and accompanying rationale.
   }\label{fig:workflow}
\end{figure}

Figure~\ref{fig:workflow} illustrates our proposed multi-agent workflow, which consists of three core agents: the \emph{therapist agent}, the \emph{client agent}, and the \emph{diagnostician agent}. 
The therapist agent initiates and conducts a structured interview with the client agent, who simulates a human client based on a predefined mental health profile. 
Upon completion of the dialogue, the full conversation transcript is passed to the diagnostician agent, which analyzes the exchange and produces a provisional diagnosis along with an explicit, step-by-step clinical rationale.
Each agent is powered by a large language model (\ac{LLM}) and guided by a carefully engineered, role-specific prompt (See Appendix~\ref{appendix:agent-prompts}) to ensure realistic and consistent behavior. 


Algorithm \ref{alg:multiagent-workflow} provides the high-level pseudocode for the multi-agent workflow:
The multi-agent mental health diagnostic workflow begins by initializing a simulated therapist, client, and diagnostician agent, each powered by a language model. 
The system tracks questionnaire items, with the therapist agent sequentially presenting each item to the client agent in a conversational format. 
The client responds according to their assigned profile, and the conversation continues until all items are addressed. 
Once the interview is complete, the diagnostician agent retrieves relevant DSM-5 passages and generates a diagnosis, including the predicted disorder, a step-by-step rationale, supporting evidence from the conversation, and a treatment recommendation. 
The process outputs the full conversation history along with the diagnosis and its explanation.
\input{tables/algorithm.tex}
In the following subsections, we describe the three agents in detail, including their design, functionality, and interaction protocols.

\subsection{Therapist Agent}
The therapist agent is responsible for administering the DSM-5 questionnaire in a conversational, adaptive manner. 
Its design combines prompt engineering, dynamic item selection, and response tracking to simulate a realistic clinical interview.
The prompt enforces strict role adherence, instructs the agent to rephrase DSM-5 items into natural, empathetic questions, and prohibits diagnostic statements or advice before the assessment is complete.

The agent follows a structured, item-tracking algorithm to ensure comprehensive coverage of all 23 DSM-5 questionnaire items:
\begin{enumerate*}
\item \textit{Initialization:}
The agent loads the list of DSM-5 questionnaire items and initializes the conversation history.
\item \textit{Iterative Questioning:}
For each item, the agent rephrases the item into a conversational question (using the LLM and the prompt template), presents it to the client agent, and appends both the question and the client’s response to the conversation history.
\item \textit{Coverage Tracking:}
After each response, the agent checks if the item has been sufficiently addressed. If not, it may rephrase or clarify the question.
\item \textit{Completion:}
The process continues until all items are addressed, ensuring full coverage of the DSM-5 domains.
\end{enumerate*}

\subsection{Client Agent}
The client agent simulates a human client with a specific mental health profile, responding to the therapist’s questions in a realistic, first-person manner. The agent is designed to maintain strict role fidelity, express symptoms according to its profile, and avoid revealing its diagnostic label.

\paragraph{Client Profile Specification}
A \textit{client profile} in our workflow is a structured description that encodes the simulated human client's mental health characteristics and contextual background.
Each client agent is initialized with a structured profile, which may include:
\begin{enumerate*}
    \item Primary disorder (e.g., Anxiety, PTSD),
    \item Comorbid modifiers: Optional secondary symptoms for increased realism,
    \item Demographics and context: Age, gender, recent life events, coping style.
\end{enumerate*}
Profiles are loaded from external files, allowing for rapid creation of diverse and edge-case personas.

\paragraph{Role Constraints and Prompt Design}
The client agent is governed by a prompt that enforces the following constraints:
\begin{enumerate*}
    \item Always respond in the first person, never as an AI or in a meta role.
    \item Never name or hint at the diagnosis.
    \item Ensure all responses are consistent with the symptoms, background, and context defined in the assigned profile.
    \item Express genuine emotional responses and appear to be seeking help.
\end{enumerate*}

During simulation, the client agent uses the assigned profile to generate consistent, first-person responses that reflect the specified symptoms and background, without explicitly naming the diagnosis.

\subsection{Diagnostician Agent}
The diagnostician agent is responsible for transforming the raw therapist–client conversation into a transparent, evidence-based diagnostic assessment. 
This agent provides a result in three distinct stages: 
\begin{enumerate*}
    \item Disorder type prediction,
    \item Rationale generation and association extraction, and 
    \item treatment recommendation.
\end{enumerate*}
Each stage is designed to maximize explainability and clinical fidelity, leveraging both \ac{RAG}~\cite{Kermani2025} and structured reasoning. 
See the demonstration in the Appendix \ref{appendix:demonstration}

\subsubsection{Disorder Type Prediction}
To ensure that diagnostic conclusions are grounded in established clinical criteria, the diagnostician agent employs a \ac{RAG} strategy. 
Specifically, the agent retrieves the top-5 most relevant DSM-5 passages for the given conversation, using a chunk size of 512/1024 tokens and the \texttt{nomic-embed-text} embedding model. This retrieval step ensures that the agent’s reasoning is explicitly linked to authoritative source material, rather than relying solely on model priors. The agent then synthesizes the retrieved information with the conversation transcript to predict the most likely disorder(s), mapping client utterances to DSM-5 diagnostic criteria. Batch processing is parallelized across four threads to accelerate data generation and evaluation.

\subsubsection{Reason Generation and Association Extraction}
Beyond producing a categorical diagnosis, the agent generates a step-by-step clinical rationale. 
This rationale explicitly cites individual client utterances as evidence for (or against) each relevant DSM-5 criterion, creating a transparent chain of reasoning. 
The agent extracts associations between symptoms mentioned in the conversation and the diagnostic criteria, highlighting both supporting and contradictory evidence. 
This process not only increases the interpretability of the diagnosis but also enables downstream auditing and error analysis.

\subsubsection{Treatment Recommendation}
Finally, the diagnostician agent provides a concise, evidence-based treatment recommendation tailored to the predicted disorder(s) and the client’s contextual profile. 
Recommendations are grounded in the retrieved DSM-5 passages and, where appropriate, reference established clinical guidelines. 
This ensures that the output is not only diagnostically accurate but also provides actionable guidance for subsequent care steps.

\section{Evaluation Protocol and Metrics}
\begin{figure}[H]
  \centering
  \includegraphics[width=0.7\linewidth]{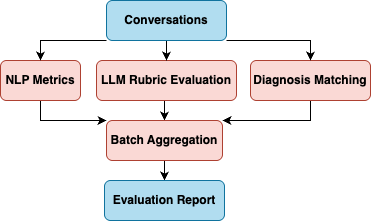}
  \caption{Evaluation protocol: Each conversation log is evaluated by the LLM using a custom rubric, NLP metrics are computed, and diagnosis accuracy is checked. Results are aggregated to produce batch-level statistics and reports.}
  \label{fig:evaluation-pipeline}
\end{figure}

 The evaluation of simulated conversations is performed using a multi-stage pipeline shown in Figure \ref{fig:evaluation-pipeline}.

\subsection{Conversation Quality Evaluation}
For evaluating the quality of the generated conversations, we employ a combination of established NLP metrics and a custom rubric-based evaluation using an LLM. 
The following metrics are computed for each conversation:
\begin{enumerate*}
    \item \textbf{Coherence}: \textit{BERTScore} \cite{zhang2020bertscoreevaluatingtextgeneration} is computed between successive conversation turns to assess logical flow or semantic coherence.
    \item \textbf{Readability}: \textit{\acf{FKG}}, \textit{\acf{GFI}} and \textit{\acf{FRE}} estimate the U.S. school grade level or reading ease of a passage using sentence length and word‐syllable counts, thereby gauging the professional register and accessibility of the conversation.
    \item \textbf{Explainability}: \acf{GFI} is also applied to the model-generated diagnostic rationales to quantify their comprehensibility and transparency.
    \item \textbf{Rubric-based LLM evaluation}: Each conversation and diagnosis is scored by an LLM using a custom rubric with five criteria (See Table ~\ref{tab:llm-rubric}).
\end{enumerate*}

\input{tables/evaluation_rubric_v2.tex}

\subsection{Diagnostic Accuracy Evaluation}
For evaluate the diagnostic accuracy of the generated conversations, we compare the predicted disorder profile against the ground-truth profile using a set of classification metrics. 
The evaluation is performed at both the individual and batch levels, with the following metrics:
\begin{enumerate*}
    \item \textbf{Classification metrics:} \textit{Precision}, \textit{Recall}, \textit{F1-Score}, and \textit{ROC-AUC} are calculated for strict matches between predicted and ground-truth DSM-5 codes.
    \item \textbf{Confusion matrix:} Analysis of systematic misclassifications.
    \item \textbf{Diagnosis accuracy:} Comparison of predicted diagnosis and the ground-truth disorder profile.
\end{enumerate*}

\section{Experimental Setup}\label{sec:experimental}

\subsection{Research Questions}
\begin{enumerate*}[label={\textbf{(RQ\arabic*)}}, nosep, leftmargin=*]
\item Can LLMs simulate therapist-client conversations, to effectively complete the diagnostic DSM-5 questionnaire, using prevailing LLMs?
\item Can disorder type predictions be made by linking questionnaire responses to specific disorder descriptions, varying prevailing LLMs?
\item Is the LLM able to make explainable and transparent diagnoses to enhance the perceived trustworthiness?
\end{enumerate*}

\subsection{Backbone \acp{LLM}}
The experimental evaluation benchmarks the proposed framework using four prevailing
LLMs: \textit{\LlamaFull}, \textit{\MistralFull}, \textit{\QwenFull} and \textit{\OpenaiFull}. 
These models were selected primarily based on their availability on the Groq cloud inference platform, as local execution was not feasible due to hardware limitations. 
The selection represents a range of open-source and proprietary architectures, balancing performance, accessibility, and diversity in language modeling approaches. 
\OpenaiShort{}, \LlamaShort{} and \MistralShort{} are models with demonstrated conversational abilities. 
\QwenShort{}, in contrast, is specifically designed as a reasoning model, which means it is optimized for step-by-step logical inference and complex decision-making, rather than purely conversational fluency. 
This distinction allows for a direct comparison between models focused on dialogue and those optimized for structured reasoning in clinical assessment tasks.


\subsection{Dataset}
A total of 8,000 simulated therapist–client conversations were generated—2,000 for each of the four LLM models. 
Each conversation is structured to include a full DSM-5 questionnaire, with the therapist agent sequentially asking questions and the client agent responding according to its assigned profile.
Client profiles are systematically diversified by primary disorders, comorbid symptoms, and demographic attributes to enhance realism and ensure broad coverage across diagnostic categories.
Simulated client profiles cover 10 primary mental disorder categories:
\begin{enumerate*}
    \item Adjustment Disorder,
    \item Anxiety,
    \item Bipolar Disorder,
    \item Depression,
    \item Obsessive-Compulsive Disorder (OCD),
    \item Panic Disorder,
    \item Post-Traumatic Stress Disorder (PTSD),
    \item Schizophrenia,
    \item Social Anxiety Disorder,
    \item Substance Abuse
\end{enumerate*}
with additional comorbidities and demographic variations.
The DSM-5 questionnaire assesses symptoms across 13 domains:
\begin{enumerate*}
    \item Depression,
    \item Anger,
    \item Mania,
    \item Anxiety,
    \item Somatic Symptoms,
    \item Suicidal Ideation,
    \item Psychosis,
    \item Sleep Problems,
    \item Memory,
    \item Repetitive Thoughts and Behaviors,
    \item Dissociation,
    \item Personality Functioning,
    \item Substance Use.
\end{enumerate*}

\subsection{Implementation Details}
The framework supports both local and cloud-based LLM inference. 
Local inference is performed using the Ollama platform, while cloud inference leverages either the Groq or OpenAI API. All conversations and diagnostic outputs were generated using the Groq and OpenAI API's.
To efficiently generate the large-scale dataset, we implemented a parallel batch processing system:
\begin{enumerate*}
    \item Workload distribution across 4 worker threads. 
    \item Exponential back off retry logic for API rate limit handling.
\end{enumerate*}
This parallelization approach reduced the total generation time from an estimated 100 hours (serial execution) to approximately 24 hours, a 4x improvement.

\section{Results}\label{sec:results}
\subsection{Performance of Conversation Quality (RQ1)}\label{sec:rq1-results}

\subsubsection{Metric-based Evaluation}

Table \ref{tab:semantic_coherence_and_readability} reports the conversation quality metrics for each model, including coherence, readability, and diversity.
\input{tables/semantic_coherence_and_readability.tex}
The conversation quality metrics indicate that all models produced dialogues with moderate global coherence (0.45--0.55 on a 0--1 scale). 
\LlamaShort{} achieved the highest \ac{FRE} score at 61.7, which is considered ``Plain English'', while \MistralShort{}, \QwenShort{} and \OpenaiShort{} had scores of 49.6, 51.1, and 53.81, respectively, which is considered ``Difficult to read''. 
The \ac{FKG} and \ac{GFI} scores ranged from 7.0 to 9.0 and 3.9 to 5.23, indicating that the text is generally readable by individuals with a middle school to early high school education.

\subsubsection{LLM Rubric Evaluation} 

\begin{figure}[h]
    \centering    \includegraphics[trim=0cm 1cm 0cm 0cm, clip, width=\columnwidth]{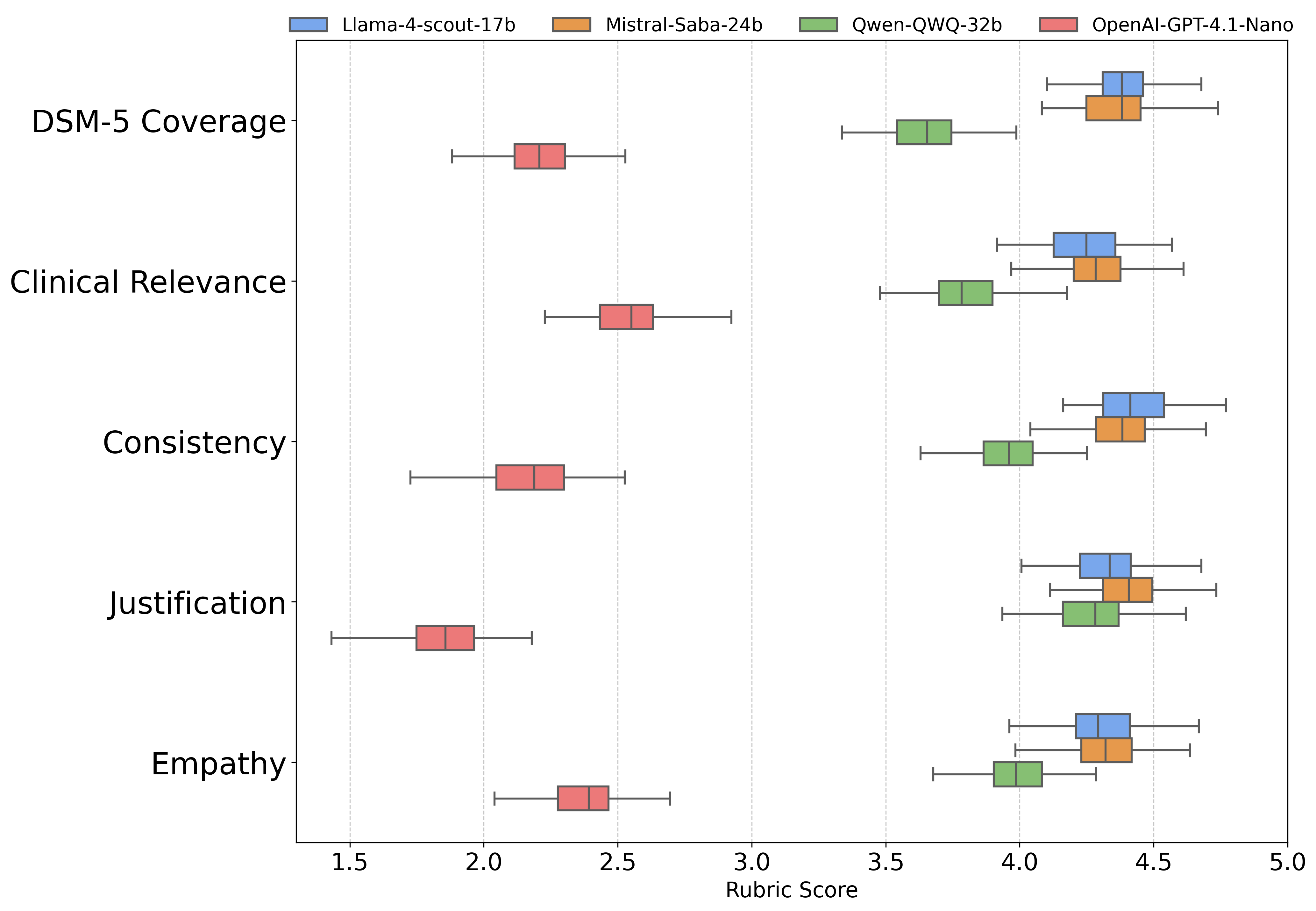}
    \caption{Boxplot of diverse LLMs' evaluation scores, ranging from 1 to 5, evaluated by multiple rubric-based metrics.}
    \label{fig:rubric-scores-barplot}
\end{figure}

Figure~\ref{fig:rubric-scores-barplot} show the boxplot of diverse LLMs' evaluation scores.
\LlamaShort{} and \MistralShort{} generally outperformed \QwenShort{}, with mean rubric scores ranging from 4.26 to 4.41 out of 5 for the top two models, compared to 3.64 to 4.23 for \QwenShort{}. 
This shows that the \QwenShort{} model performs approximately 9.23\% worse. The top two models performed similarly throughout all categories.
\OpenaiShort{}, on the other hand, performs substantially worse compared to the rest. With mean rubric scores ranging from 1.89 to 2.54, this indicates a performance drop of approximately 48.91\% relative to the top models.

\subsection{Performance of Disorder Diagnosis (RQ2)}\label{sec:rq2-results}

\subsubsection{Overall Diagnostic Performance Across Models}

\begin{figure}[H]
    \centering
    \includegraphics[trim=0cm 1cm 0cm 0cm, clip, width=0.95\columnwidth]{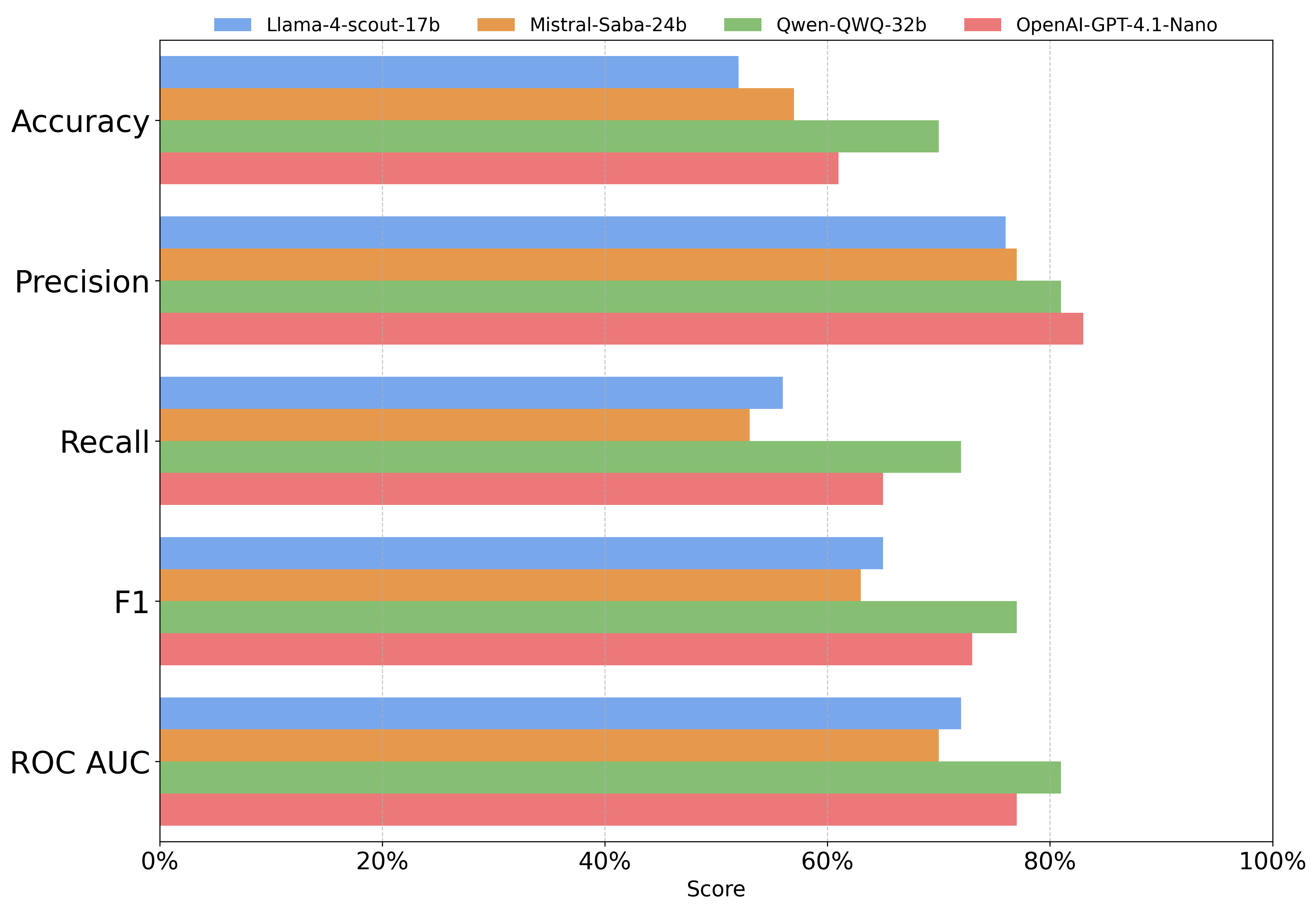}
    \caption{Diagnostic classification performance for diverse LLMs, evaluated by F1 score (\%).}
    \label{fig:diagnostic-metrics-multimodel-barplot}
\end{figure}

As shown in Figure~\ref{fig:diagnostic-metrics-multimodel-barplot}, the diagnostic accuracy metrics reveal that \QwenShort{} achieved the highest overall performance, notably outperforming the other models with an accuracy of 70\%, recall of 72\%, and an F1 score of 77\%. \OpenaiShort{} followed with strong results across all metrics, achieving an F1 score of 73\% and the highest precision at 83\%. \LlamaShort{} and \MistralShort{} showed relatively lower performance, with accuracies of 52\% and 57\%, and F1 scores of 65\% and 63\%, respectively.

\subsubsection{Diagnosis Performance Across Types and Confusion Matrix}
\input{tables/f1_score.tex}

As shown in Table~\ref{tab:perlabelf1}, the per-label F1 scores indicate that all models performed best on anxiety-related disorders—namely Anxiety, Panic, PTSD, and Social Anxiety—with F1 scores above 80\% in most cases. Notably, \QwenShort{} achieved over 93\% on Panic (93.65\%), PTSD (94.36\%), and social anxiety (93.89\%), while \OpenaiShort{} surpassed all models for PTSD with 98.53\% and performed competitively on OCD (95.81\%) and Bipolar disorder (95.28\%). In contrast, performance was substantially lower on Adjustment Disorder, with \LlamaShort{}, \MistralShort{}, and \OpenaiShort{} scoring below 3\%, and only \QwenShort{} reaching 40.25\%. Depression also posed challenges, with scores ranging from 36.75\% to 67.98\%. Overall, \QwenShort{} consistently achieved top performance across most diagnostic categories, closely followed by \OpenaiShort{}, which showed strong results in several high-complexity disorders.

\begin{figure*}[t]
    \centering
    \begin{subfigure}[t]{0.255\textwidth}
        \centering
        \includegraphics[trim=0.8cm 0.8cm 0cm 0cm, clip, width=\linewidth]{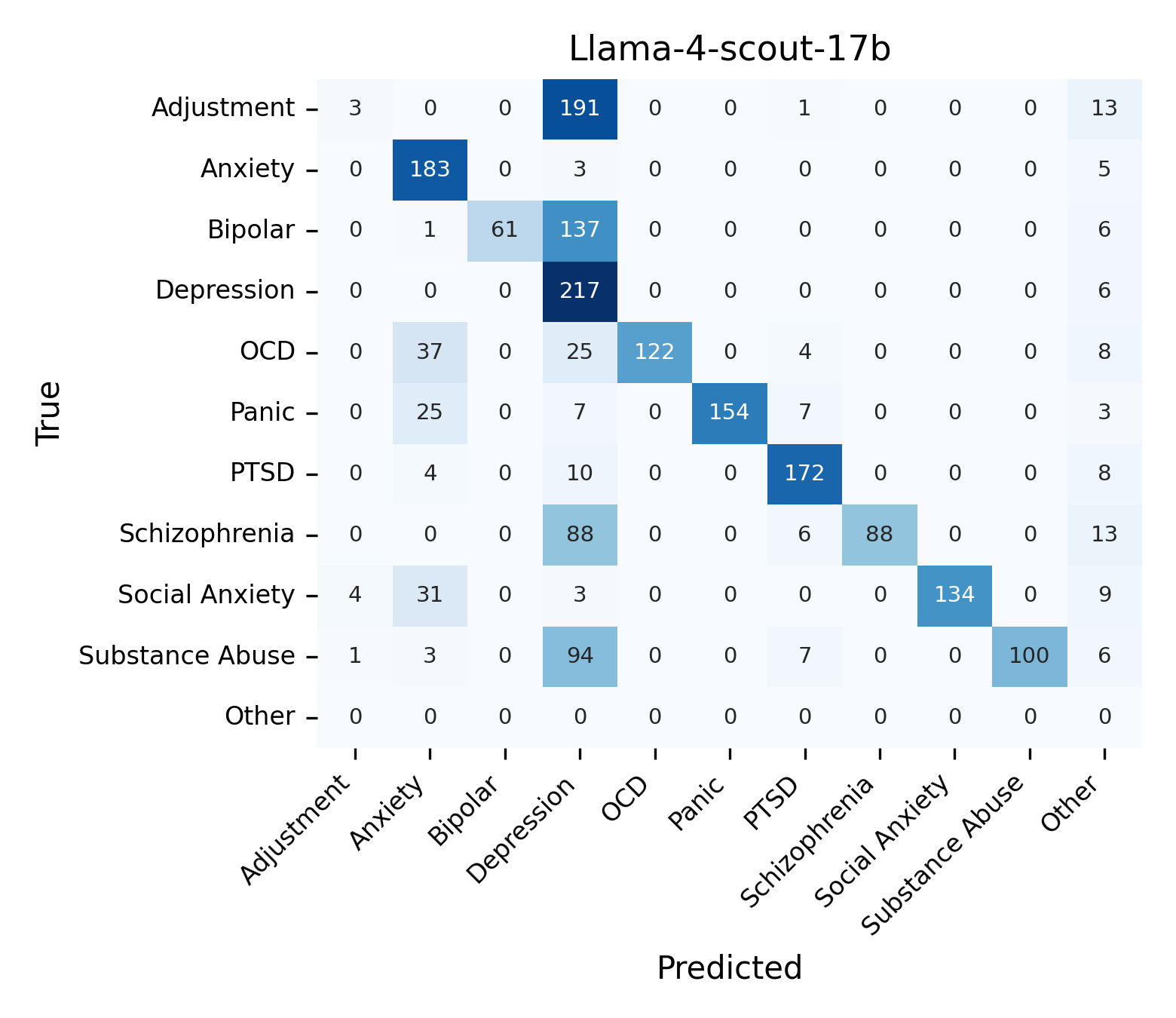}
    \end{subfigure}
    \hfill
    \begin{subfigure}[t]{0.24\textwidth}
        \centering
        \includegraphics[trim=0.5cm 0cm 0cm 0cm, clip, width=\linewidth]{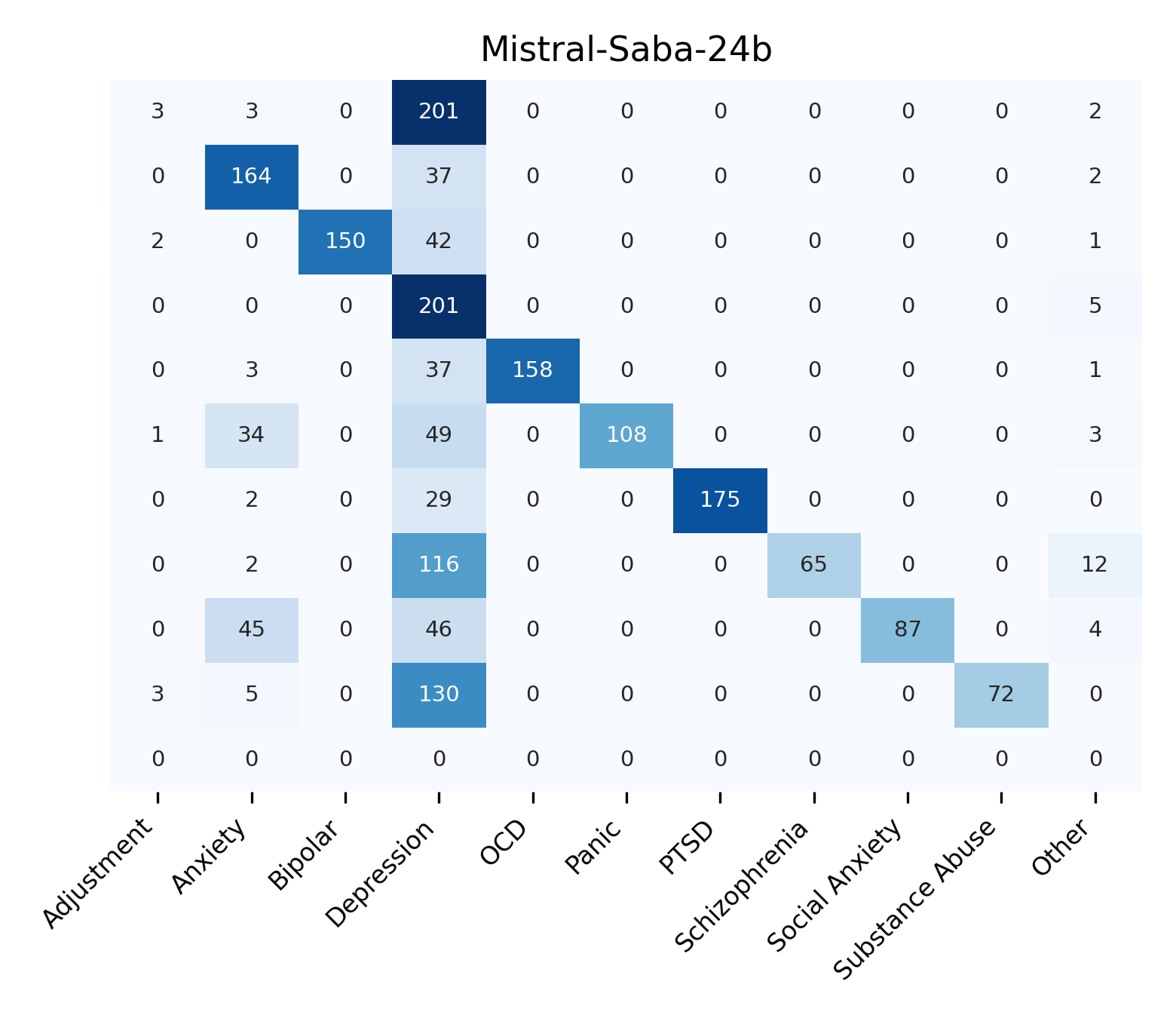}
    \end{subfigure}
    \hfill
    \begin{subfigure}[t]{0.24\textwidth}
        \centering
        \includegraphics[trim=0.5cm 0cm 0cm 0cm, clip,width=\linewidth]{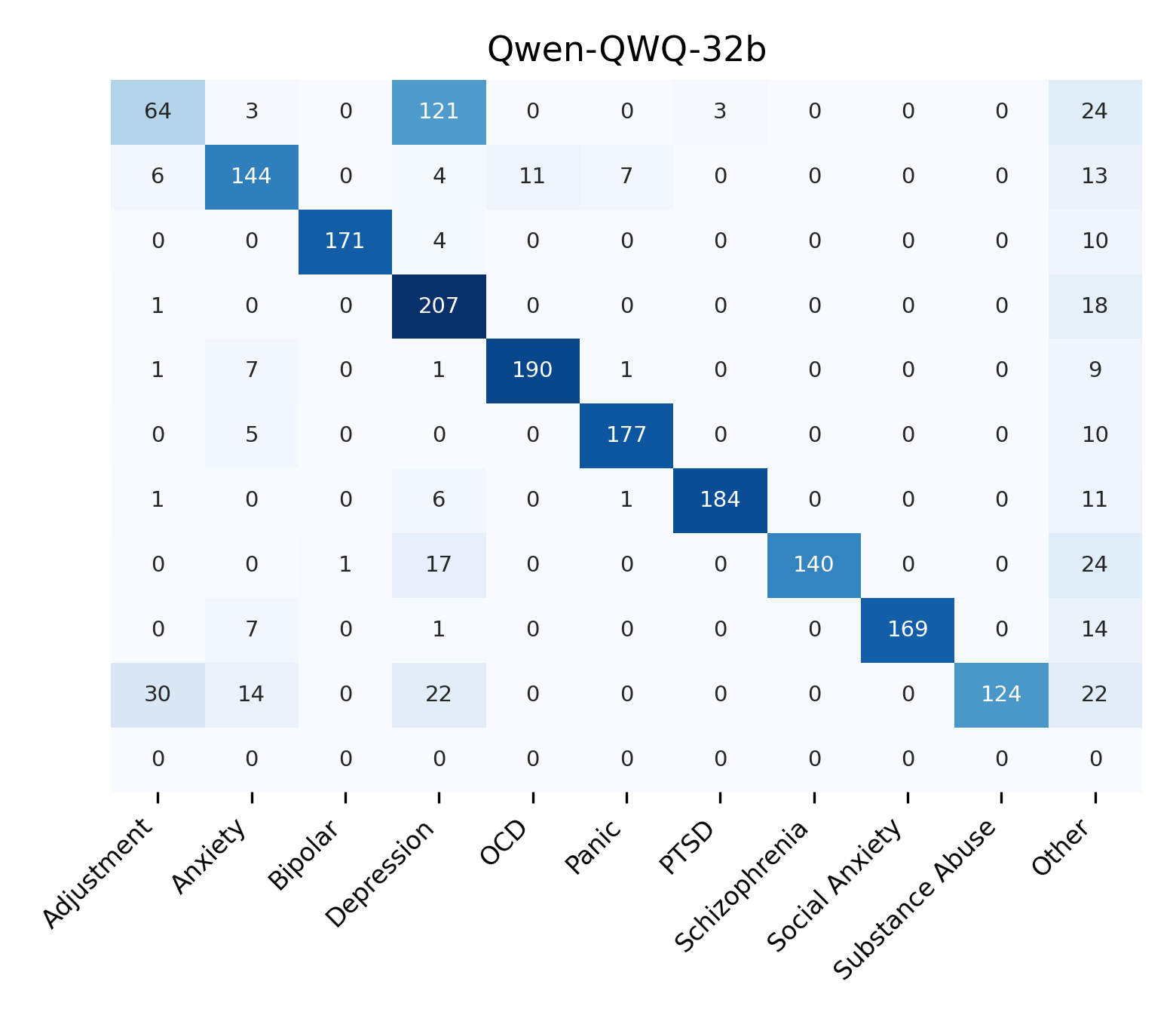}
    \end{subfigure}
    \hfill
    \begin{subfigure}[t]{0.24\textwidth}
        \centering
        \includegraphics[trim=0.5cm 0cm 0cm 0cm, clip,width=\linewidth]{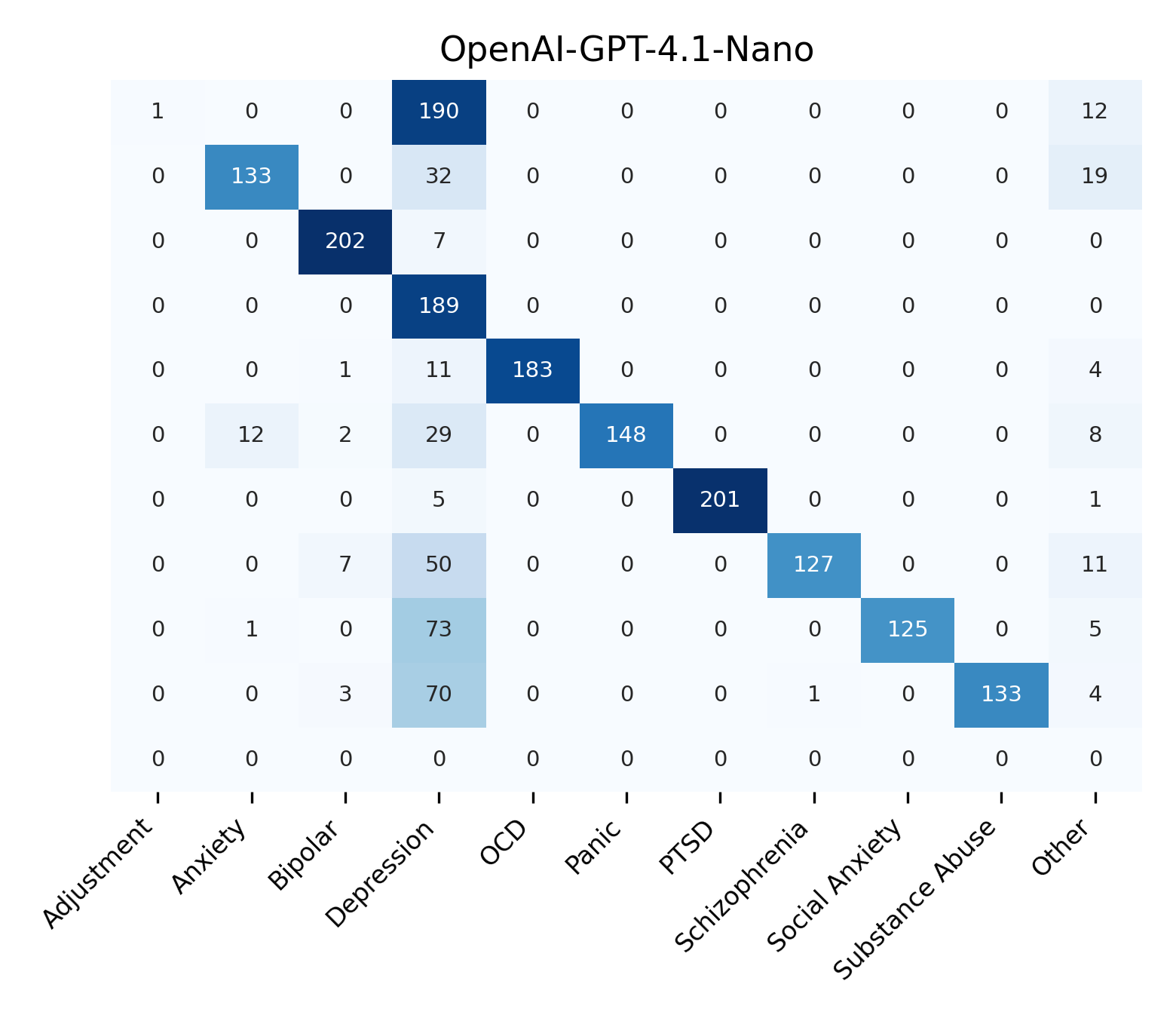}
    \end{subfigure}
    \caption{Confusion matrices for all LLMs across all diagnostic profiles. Each matrix shows the distribution of predicted (x-axis) versus ground-truth (y-axis) disorder labels, highlighting systematic misclassifications.}   \label{fig:confusion-matrix}
\end{figure*}

The confusion matrices (Figure~\ref{fig:confusion-matrix}) reveal consistent misclassification patterns across all models. Adjustment Disorder is frequently mislabeled as Depression, especially by \LlamaShort{}, \MistralShort{}, and \OpenaiShort{}. While \QwenShort{} performs better, it still misclassifies many Adjustment cases. Bipolar disorder and Depression are often confused.

Social Anxiety and Substance Abuse also show confusion, often being mislabeled as Anxiety and Depression, respectively. Despite these issues, all models perform well on Anxiety, Panic, PTSD, and OCD, with high correct classification rates. \QwenShort{} and \OpenaiShort{} show the most consistent performance, though challenges remain for disorders with overlapping symptoms.

\subsection{Case Study for Explainability (RQ3)}\label{sec:rq3-results}

A practical goal of the multi-agent design is not merely to reach the
\emph{right} diagnosis but to expose the \emph{reasoning chain} in a
way clinicians can audit.  We therefore inspected one randomly chosen transcript per backbone and scored the Diagnostician’s answer on three transparency signals:
\begin{enumerate*}[leftmargin=*, nosep]
  \item \textit{Evidence tags} — frequency of \texttt{<sym>}, \texttt{<quote>}, \texttt{<med>} tags that explicitly link utterances to criteria.
  \item \textit{Criterion anchors} — whether the text cites DSM-5 clauses (``criterion A'', ``criterion C'', etc.).
  \item \textit{Step-by-step logic} — presence of numbered or bullet reasoning steps that map evidence\,$\rightarrow$\,conclusion.
\end{enumerate*}
We report the statistics of explainability signals in Table~\ref{tab:explain-matrix}.
\input{tables/explain_matrix.tex}

\textit{Qwen-QWQ-32b -- an ideal explanation following the standard criteria\footnote{\url{https://www.ncbi.nlm.nih.gov/books/NBK519704/table/ch3.t22/}}.}
The Diagnostician produced a five-point, numbered rationale:
\begin{enumerate*}
    \item \texttt{<sym>}Auditory hallucinations for 3 weeks\texttt{</sym>} satisfy DSM-5 \textit{Criterion A1}.
    \item No mood episode longer than the psychosis (\textit{criterion D}).
    \item Normal toxicology panel excludes substance aetiology (\textit{Criterion E}).
    \item \texttt{<quote>}``I know these voices aren’t real''\texttt{</quote>} indicates partial insight.
    \item Hence provisional \texttt{<med>}Schizophreniform Disorder\texttt{</med>}.
\end{enumerate*}
Eleven symptom tags and four direct client quotes make the evidence
traceable; each sentence maps to a clause, giving reviewers a clear
audit trail.

\textit{Mistral-Saba-24b -- accurate but harder to audit.}
Although the primary diagnosis (\emph{Social Anxiety}) is
correct, the rationale is a single paragraph paraphrasing the criteria.
Only two client quotes appear, and there is no numbered logic, forcing a reader to infer which symptom supports which clause.

\textit{Llama-4-Scout-17b -- opaque black-box output.}
The model lists three disorders—
\texttt{<med>}MDD, GAD, Panic\texttt{</med>}—with minimal justification:
``Symptoms appear consistent with DSM-5 mood and anxiety criteria.''
No criteria are cited; no client quotes are surfaced.  The result
achieves the lowest explainability by every metric.

\textit{GPT-4.1-Nano -- rich tags but weak structure.}
It inserts 29 symptom tags and 6
\texttt{\textless med\textgreater} tags -- more evidence markers than the others -- yet never references DSM clauses and offers no
step-wise logic.

\section{Related Work}\label{sec:related-work}

\subsection{Mental Health Detection and Diagnosis}
LLMs are increasingly used for screening and diagnosing conditions like depression and anxiety from text data \cite{DeDuro2025, Kermani2025}. Initiatives like MDD-5k~\cite{Yin2024} and CounseLLMe~\cite{DeDuro2025} demonstrate the use of LLM-generated dialogues for dataset creation. Recent reviews and studies indicate LLMs can match or even exceed human performance in some diagnostic tasks~\cite{Hua2025, levkovich2025evaluating}, and their utility is expanding into broader psychiatric research~\cite{LLMsPsychiatricResearchPMC2025}. Methods vary, with fine-tuning often yielding high accuracy, \ac{RAG} helping to ground responses~\cite{Kermani2025}, and novel approaches like translating diagnostic manuals into inspectable logic programs for interpretable diagnoses emerging~\cite{Kim2025}. Some work focuses on converting unstructured interviews to structured data for assessment \cite{LLMQuestionnaireCompletionACL2024}. 
Despite progress, challenges like bias and the need for robust reasoning persist, guiding best practices towards multi-agent systems and explainable rationales \cite{ChenLiu2025, Kim2025}. 

Recent work~\cite{NaturalLanguageMedicationRec} shows that grounding LLM reasoning in \emph{structured} and \emph{multi-modal} client information can further improve clinical decision-making fidelity. 
\acs{NLA-MMR}~\cite{NaturalLanguageMedicationRec} aligned the embeddings of drug-description with tabular EHR features to produce personalized treatment suggestions. 
Complementing this, \emph{CausalMed}~\cite{CausalMed} frames medication choice as counterfactual reasoning over client–disease–drug causal graphs, achieving superior performance to association-based baselines. 
While these studies target medication recommendation rather than screening, their architectures -- multi-modal fusion and causality-aware reasoning -- inform our design goal of combining conversational signals with DSM-5 knowledge to yield transparent diagnostic outcomes.
Several works~\citep{sun2024eliciting,sun2025rethinking,sun2024script} explored the alignment of psychotherapy dialogues with psychological strategies.

While existing research demonstrates the growing capability of LLMs in mental health screening using various datasets and analytical techniques, our study introduces a novel multi-agent workflow specifically designed to automate the completion of DSM-5 questionnaires through simulated dialogues. The innovation lies not just in detection, but in the structured generation of these dialogues by distinct therapist and client agents, followed by a dedicated diagnostician agent that provides transparent disorder predictions. This system architecture directly addresses the challenge of opaque diagnostic processes by making the reasoning explicit and grounded in DSM-5 criteria, a more granular approach than general text analysis for symptom identification.

\subsection{Synthetic Conversations and Explainability}
Generating synthetic mental health dialogues is a key strategy to overcome data scarcity while aiming for transparency \cite{Guo2024, Stade2024}. Advanced generative agents can produce clinically relevant conversations, though oversight is crucial \cite{Guo2024, BenZion2025}. Frameworks such as DiaSynth employ \ac{CoT} reasoning for dynamic dialogue generation \cite{DiaSynthACL2025}, and LLM-generated data has been used to balance datasets for tasks like depression detection and improve empathy recognition models \cite{SyntheticDataDepressionArXiv2024, ArtificialPsychotherapyDialoguesACL2025}. The fidelity of these synthetic conversations is also an active area of research \cite{SyntheticFidelityPTSDArXiv2025}.

Explainability is vital for trust in diagnostic systems. 
Step-by-step reasoning enhances perceptions of fairness and reliability~\cite{Xu2024}, aligning with the broader push for \ac{XAI} in mental health~\cite{Kim2025}. 
Barkan et al.~\cite{LearningExplainableModels} propose a learning-based attribution model that generates concise, human-interpretable explanations for language-model predictions and outperforms attention heat-maps on faithfulness metrics. 
Their findings reinforce our choice to embed a dedicated diagnostician agent that produces intrinsic rationales rather than relying solely on post-hoc attention visualisations. LLMs themselves are being explored to make complex AI outputs more understandable~\cite{LLMsForXAISurveyArXiv2025, XAIHealthcareTrendsMedRxiv2024}, and ensuring the credibility and ethical use of LLM-based mental health tools is a major research focus~\cite{CredibilityLLMsResearchProtocols2025}.

The innovative aspect of this study, compared to the existing work on synthetic conversation generation~\cite{pei2025conversational} and explainability~\cite{LearningExplainableModels}, is its integrated multi-agent system, where synthetic dialogues are not only generated but are intrinsically linked to an explainable diagnostic process. While other studies focus on generating diverse synthetic data or on post-hoc explainability methods, this paper's framework embeds explainability within the data generation and diagnostic pipeline itself. The diagnostician agent's role in retrieving relevant DSM-5 passages and explicitly linking conversational utterances to diagnostic criteria provides a transparent, self-documenting rationale. This moves beyond general synthetic dialogue by creating conversations specifically designed to populate a standardized diagnostic tool (e.g., DSM-5) and immediately subjecting them to an interpretable analysis, thus tackling both data scarcity and the ``black box'' problem in a unified manner.

\section{Discussion}
\textbf{Implications and Future Work.}
LLM-based systems offer promise for augmenting mental health services, but ethical considerations regarding autonomy, misuse, and liability are paramount \cite{Stade2024}. Current text-based LLMs also miss nonverbal cues. Future work requires robust data curation, domain-specific fine-tuning, and interdisciplinary collaboration. The success of multi-modal and causality-aware systems such as NLA-MMR and CausalMed \cite{NaturalLanguageMedicationRec, CausalMed} also suggests that future mental-health models will benefit from integrating questionnaire text with physiological, imaging, or behavioural signals in a causally coherent manner. Multi-agent LLM systems are also showing potential in complex healthcare scenarios, such as mitigating cognitive biases in diagnosis \cite{MultiAgentCognitiveBiasJMIR2024} or simulating evolving clinical interactions \cite{SelfEvolvingMultiAgentArXiv2025}, though system complexity and cost remain considerations.
The research also highlighted persistent difficulties in accurately classifying certain disorders (e.g., Adjustment, Depression) across all models using simulated data. This may be due to substantial symptom overlap between these disorders in the DSM-5 questionnaire, which lacks the contextual and temporal detail needed for fine-grained differentiation. As a result, both LLMs and human raters may struggle to distinguish these conditions based solely on high-level screening data, a limitation that underscores the need for more granular assessment tools. 

This paper introduces a transparent, multi-agent LLM workflow for simulating and interpreting DSM-5 diagnostic interviews. While this represents only one aspect of mental health diagnosis, we believe that our core approach -- simulating therapist–client dialogues for specific client profiles and employing a multi-agent, fine-grained workflow -- offers valuable insights for the development of LLM-based agent systems in broader, real-world mental health applications. We hope this work inspires further research into multi-role architectures and their potential to enhance transparency, realism, and effectiveness in AI-driven mental health support.

\textbf{Limitations.}\label{sec:limitations}
While promising, our investigation has several constraints:
\begin{enumerate*}
\item \textit{Simulated data only.}  No real client transcripts or clinician interactions were used; ecological validity must be confirmed with human subjects. This is a critical limitation, as clinical conversations are the gold standard for evaluating diagnostic dialogue systems. While our synthetic data is designed to mirror real consultations using DSM-5 criteria and naturalistic phrasing, we cannot yet quantify how closely it approximates genuine clinical discourse. 
\item \textit{One‑shot generation.}  To reduce GPU cost, conversations were produced in a single forward pass. This precludes true turn‑by‑turn adaptation and may inflate coherence metrics.
\item \textit{Automatic evaluation bias.}  Rubric scoring relied on another LLM, which could share error modes or biases with the models under test.
\item \textit{Limited model pool.} Hardware availability constrained us to several Groq‑hosted backbones and one OpenAI backbone; larger or clinically fine‑tuned models remain unexplored.
\end{enumerate*}

\textbf{Framework Modularity and Extensibility.}\label{sec:modularity}
Our framework was engineered specifically for the mental health domain and to be \emph{plug–and–play}, enabling rapid replication and extension by other researchers:
\begin{enumerate*}
\item \textit{Backend abstraction.} 
All language generation calls are routed through a thin adapter layer that conforms to the \texttt{chat/completions} API schema. 
When the \texttt{ollama} flag is enabled, requests are served locally via Ollama, allowing open-source GGUF or HuggingFace model to be used without vendor lock–in or cloud egress of sensitive data. 
\item \textit{Custom assessment instruments.} Questionnaires are loaded at run time from PDF/TXT/Markdown files that specify section headings, item text, and scoring rules. New or proprietary mental health instruments can therefore be incorporated by adding a new file.
\item \textit{Persona definitions.} Client profiles -- including demographic attributes, diagnostic priors, and conversation goals -- are expressed in compact TXT files. 
Researchers can craft edge–case personas or under-served demographic groups by editing a few key–value pairs.
\end{enumerate*}

\textbf{Ethical Considerations.}
This study uses \emph{only} synthetic data generated by \ac{LLM} agents under researcher control; no real client data or identifiers were involved, and thus IRB approval was not required. 
\textit{Respect for persons.} Simulated clients were based on public diagnostic criteria and prevalence data, with prompts restricting any content resembling real individuals. Released data prohibit re-identification or clinical use without separate ethical approval. 
\textit{Beneficence and non-maleficence.} The system is strictly for research into explainable AI. It is \emph{not} a medical device, and its outputs must not inform clinical decisions. Misclassification risks (\S~\ref{sec:rq2-results}) underscore potential harms if misused.

\section{Conclusion}\label{sec:conclusion}

This study addressed the challenge of opacity in initial mental health screening by proposing and evaluating a novel multi-agent LLM workflow. 
The objective was to simulate DSM-5 questionnaire-related dialogues, automate the generation of transparent, step-by-step diagnostic predictions grounded in clinical criteria, and create a large-scale synthetic dataset for research.

The successful study demonstrated the feasibility of this multi-agent approach. 
The framework consistently generated coherent dialogues covering all DSM-5 domains and produced diagnostic rationales linked to conversational evidence. 
Comparative benchmarking of four distinct LLMs revealed a key trade-off: 
models optimized for conversation yielded higher scores on dialogue quality metrics, 
while a model designed for reasoning (\QwenShort{}) achieved superior diagnostic accuracy, particularly for diagnostically challenging conditions like Adjustment Disorder and Bipolar Disorder, despite lower conversational ratings.

The main finding is that multi-agent LLM systems offer a viable pathway towards creating more transparent and accountable AI-assisted tools for mental health screening. 
By explicitly simulating the diagnostic reasoning process and linking conclusions back to conversational evidence and DSM-5 criteria, the workflow moves beyond opaque predictions. 
Furthermore, the methodology offers a promising foundation for benchmarking LLM performance on clinical tasks and for generating privacy-preserving synthetic data that may contribute to future research. 
While this study demonstrates encouraging initial results, the use of simulated dialogues introduces inherent limitations. 
Further validation, including expert human review and targeted assessment of diagnostic rationale quality, is needed. 
Despite these limitations, our work provides a foundation for future systems that aim to improve transparency and support clinical decision-making in mental health screening.

\newpage
\section*{GenAI Usage Disclosure}
\Ac{LLM} tools were employed in two distinct phases of this project:

\begin{itemize}
  \item \textbf{Experimental pipeline.}  Four inference-only models—\emph{\LlamaFull}, \emph{\MistralFull}, \emph{\QwenFull} and \emph{\OpenaiFull}—were used as (i) simulated therapists, (ii) simulated clients, and (iii) diagnostician agents (§\ref{sec:method}).  Their raw outputs (dialogues, rationales, and scores) are reported verbatim except where redacted for length.
  \item \textbf{Author assistance.}  A commercial chat-based LLM was consulted sparingly for code-refactoring suggestions. All such suggestions were manually reviewed and, where necessary, edited for accuracy and style.  No figures, tables, metrics, or conclusions were generated automatically.
\end{itemize}

The authors affirm that all analyses, statistical computations, and visualisations were executed with conventional programming tools (Python, Pandas, scikit-learn, Matplotlib); 
LLMs were \emph{not} employed to fabricate data or to conduct quantitative evaluation. 
The manuscript’s final content remains the sole responsibility of the human authors.
\bibliographystyle{splncs04}
\bibliography{references}

\appendix
\onecolumn
\section{Appendix}
\subsection{Demonstration: Conversational Generation and Chat Interface}\label{appendix:demonstration}
\begin{figure*}[htb!]
  \centering
  \includegraphics[trim=0.25cm 2.6cm 0.5cm 0.5cm, clip,, width=0.85\linewidth]{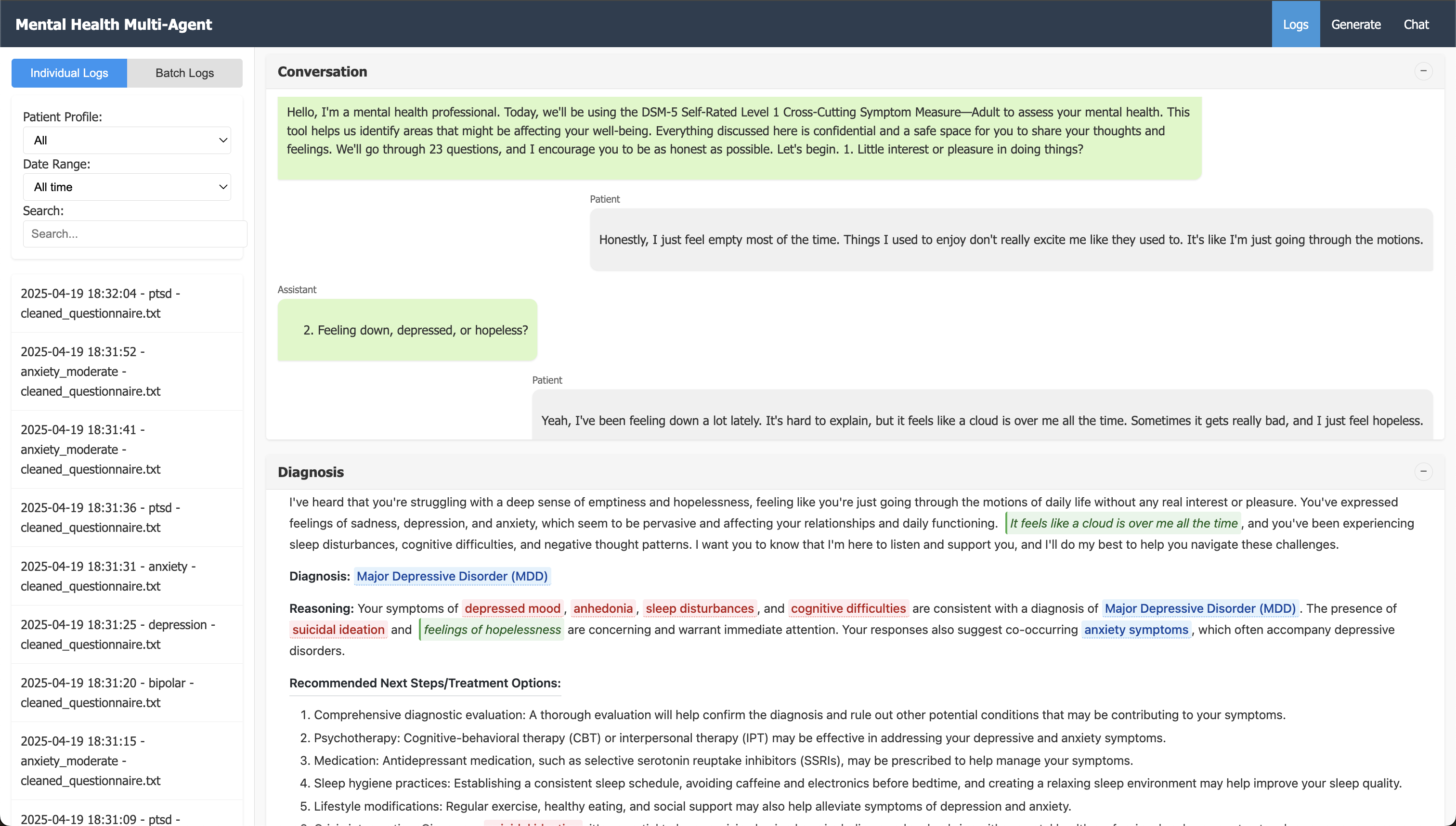}
  \caption{Demo of the conversation logs page, showing how a diagnosis and conversation look.}
  \label{fig:multi-agent-framework}
\end{figure*}

\subsection{Agent Prompts}\label{appendix:agent-prompts}
\input{tables/prompt_template.tex}

\end{document}

%% file: _package.tex
\usepackage{subcaption}
\usepackage[inline]{enumitem} 
\setlist[itemize]{leftmargin=*, nosep}
\usepackage{multirow}
\usepackage{algorithm}
\usepackage{algorithmic}
\usepackage{xcolor}

\usepackage[inline]{enumitem} 
\setlist[itemize]{leftmargin=*, nosep}

%% file: _definition.tex
\usepackage{acronym}
\acrodef{AI}{artificial intelligence}
\acrodef{LLM}{large language model}
\acrodef{FKG}{flesch-kincaid grade}
\acrodef{GFI}{gunning fog index}
\acrodef{FRE}{flesch reading ease}
\acrodef{RAG}{retrieval augmented generation}
\acrodef{NLA-MMR}{natural language-assisted multi-modal medication recommendation}
\acrodef{CoT}{Chain-of-Thought}
\acrodef{XAI}{explainable AI}
\acrodef{SOTA}{state-of-the-art}

\newcommand{\LlamaFull}{Llama-4-scout-17b\footnote{\url{https://huggingface.co/meta-llama/Llama-4-Scout-17B-16E}}}
\newcommand{\MistralFull}{Mistral-Saba-24b\footnote{\url{https://mistral.ai/news/mistral-saba}}}
\newcommand{\QwenFull}{Qwen-QWQ-32b\footnote{\url{https://huggingface.co/Qwen/QwQ-32B}}}
\newcommand{\OpenaiFull}{OpenAI GPT-4.1-Nano\footnote{\url{https://openai.com/index/gpt-4-1/}}}
\newcommand{\LlamaShort}{Llama-4}
\newcommand{\MistralShort}{Mistral-Saba}
\newcommand{\QwenShort}{Qwen-QWQ}
\newcommand{\OpenaiShort}{GPT-4.1-Nano}

\newcommand{\cmark}{\ensuremath{\checkmark}}   
\newcommand{\xmark}{\ensuremath{\times}}       

\newcommand{\OurModel}[1]{DSM5AgentFlow}

%% file: tables/algorithm.tex
\begin{algorithm}[t]
\caption{Multi-Agent Mental Health Diagnostic Workflow}
\label{alg:multiagent-workflow}
\begin{algorithmic}
\STATE \textbf{procedure} SimulateDiagnosticSession($client, model, quest$)
    \STATE $convHistory \leftarrow$ InitializeConversation()
    \STATE $therapist \leftarrow$ InitializeTherapistAgent($model, quest$)
    \STATE $client \leftarrow$ InitializeClientAgent($model, client$)
    \STATE $diagAgent \leftarrow$ InitializeDiagnosticianAgent($model$)
    
    \STATE // Initialize questionnaire tracking
    \STATE $completed \leftarrow \emptyset$
    \STATE $pending \leftarrow$ GetAllItems($quest$)
    
    \WHILE{$pending \neq \emptyset$}
        \STATE // Therapist turn
        \STATE $item \leftarrow$ SelectNextItem($pending, completed$)
        \STATE $tPrompt \leftarrow$ FormatTherapistPrompt($convHistory, item$)
        \STATE $tMessage \leftarrow$ GenerateMessage($model, tPrompt$)
        \STATE $convHistory$.Append($tMessage$)
        
        \STATE // Client turn
        \STATE $cPrompt \leftarrow$ FormatClientPrompt($convHistory, client$)
        \STATE $cMessage \leftarrow$ GenerateMessage($model, cPrompt$)
        \STATE $convHistory$.Append($cMessage$)
        
        \STATE // Update tracking
        \IF{IsItemAddressed($cMessage, item$)}
            \STATE $pending$.Remove($item$)
            \STATE $completed$.Add($item$)
        \ENDIF
    \ENDWHILE
    
    \STATE // Diagnosis generation
    \STATE $dsmPassages \leftarrow$ RetrieveDSM5($convHistory$)
    \STATE $diagPrompt \leftarrow$ FormatPrompt($convHistory, dsmPassages$)
    \STATE $diagnosis \leftarrow$ GenerateDiagnosis($model, diagPrompt$)
    
    \STATE // Extract components
    \STATE $disorder \leftarrow$ ExtractDisorder($diagnosis$)
    \STATE $rationale \leftarrow$ ExtractRationale($diagnosis$)
    \STATE $evidence \leftarrow$ ExtractEvidence($diagnosis, convHistory$)
    \STATE $treatment \leftarrow$ ExtractTreatment($diagnosis$)
    
    \STATE \textbf{return} $\{convHistory, disorder, rationale, evidence, treatment\}$
\STATE \textbf{end procedure}
\end{algorithmic}
\end{algorithm}

%% file: tables/evaluation_rubric_v2.tex
\begin{table}[h!]
\footnotesize
\caption{LLM evaluation rubric for therapist–client conversations (5 = best, 1 = worst).}
\label{tab:llm-rubric}
\resizebox{\columnwidth}{!}{
\begin{tabular}{@{}p{\linewidth}@{}}
\toprule
\textbf{Criterion / Scoring Guidelines (1–5)} \\
\midrule
\textbf{Completeness of DSM‑5 Dimension Coverage} \\
1 – DSM‑5 dimensions are barely addressed or completely ignored. \\
2 – Few DSM‑5 dimensions are explored, leading to significant gaps. \\
3 – Some DSM‑5 dimensions are missing or only superficially covered. \\
4 – Most DSM‑5 dimensions are addressed, with minor omissions. \\
5 – All relevant DSM‑5 Level 1 dimensions are thoroughly explored. \\
\midrule
\textbf{Clinical Relevance and Accuracy of Questions} \\
1 – Questions are unrelated or inappropriate for clinical assessment. \\
2 – Questions poorly reflect DSM‑5 criteria; most clinically irrelevant. \\
3 – Questions somewhat reflect DSM‑5 criteria but several inaccuracies exist. \\
4 – Questions generally align with DSM‑5 criteria with slight inaccuracies. \\
5 – Questions precisely reflect DSM‑5 criteria, clearly targeting symptoms. \\
\midrule
\textbf{Consistency and Logical Flow} \\
1 – Dialogue appears random or highly disconnected, lacking progression. \\
2 – Frequent inconsistencies severely impact coherence. \\
3 – Noticeable inconsistencies occasionally disrupt understanding. \\
4 – Minor inconsistencies exist but do not significantly disrupt flow. \\
5 – Dialogue flows logically; each question naturally follows responses. \\
\midrule
\textbf{Diagnostic Justification and Explainability} \\
1 – Diagnoses have no clear connection to content or lack justification. \\
2 – Diagnoses rarely align with responses; justifications are superficial or unclear. \\
3 – Diagnoses somewhat align but have partially flawed justifications. \\
4 – Diagnoses generally align with responses, minor ambiguity in rationale. \\
5 – Diagnoses clearly align with DSM‑5 responses, reasoning explicitly stated. \\
\midrule
\textbf{Empathy, Naturalness, and Professionalism} \\
1 – Completely lacking empathy, professional tone, or natural flow. \\
2 – Rarely empathetic; responses generally impersonal, abrupt, or inappropriate. \\
3 – Occasional empathy; interactions are sometimes robotic or impersonal. \\
4 – Mostly empathetic and professional, with minor unnatural moments. \\
5 – Consistently empathetic, natural conversational style, and professional tone. \\
\bottomrule
\end{tabular}
}
\end{table}

%% file: tables/semantic_coherence_and_readability.tex
\begin{table}[t]
\centering
\footnotesize
\caption{Conversation-quality evaluation on semantic coherence and readability across multiple \ac{LLM} backbones. The symbol $\uparrow$ indicates higher is better and and $\downarrow$ indicate lower is better. }
\label{tab:semantic_coherence_and_readability}
\resizebox{\columnwidth}{!}{
\begin{tabular}{@{}cccccc@{}}
\toprule
\multirow{2}{*}{\textbf{LLM}} 
 & \multicolumn{1}{c}{\textbf{Coherence}} 
 & \multicolumn{1}{c}{}
 & \multicolumn{3}{c}{\textbf{Readability}} \\ 
\cmidrule(lr){2-3} \cmidrule(lr){4-6}
 & \textbf{BERTScore $\uparrow$} &
 & \textbf{\ac{FRE} $\uparrow$} 
 & \textbf{\ac{FKG} $\downarrow$} & \textbf{\ac{GFI}$\downarrow$} \\ \midrule
Llama-4-scout-17b & 50.77 \% & & 61.67 & 7.01 & 3.87 \\
Mistral-Saba-24b   & 51.30 \% &  & 49.58 & 8.99 & 4.35 \\
Qwen-QWQ-32b       & 50.68 \% & & 51.10 & 8.70 & 4.21 \\
GPT-4.1-Nano       & 54.87 \% & & 53.81 & 8.96 & 5.23 \\
\bottomrule
\end{tabular}}
\end{table}

%% file: tables/f1_score.tex
\begin{table}
\centering
\footnotesize
\caption{Diagnosis performance across disorder types for diverse LLMs, evaluated by F1 score (\%).}
\label{tab:perlabelf1}
\resizebox{\columnwidth}{!}{
\begin{tabular}{lcccc}
\hline
\textbf{Disorder Type} & \textbf{\LlamaShort} & \textbf{\MistralShort} & \textbf{\QwenShort} & \textbf{\OpenaiShort} \\
\hline
Adjustment & 2.78\% & 2.75\% & 40.25\% & 0.98\% \\
Anxiety & 88.10\% & 81.10\% & 88.28\% & 78.60\% \\
Bipolar & 45.86\% & 86.96\% & 95.80\% & 95.28\% \\
Depression & 43.49\% & 36.75\% & 67.98\% & 44.73\% \\
OCD & 76.73\% & 88.52\% & 92.68\% & 95.81\% \\
Panic & 88.00\% & 71.29\% & 93.65\% & 85.30\% \\
PTSD & 87.98\% & 91.86\% & 94.36\% & 98.53\% \\
Schizophrenia & 62.19\% & 50.00\% & 86.96\% & 78.64\% \\
Social Anxiety & 85.08\% & 64.68\% & 93.89\% & 75.99\% \\
Substance Abuse & 64.31\% & 51.06\% & 73.81\% & 77.33\% \\
\hline
\end{tabular}}
\end{table}

%% file: tables/explain_matrix.tex
\begingroup
\renewcommand{\arraystretch}{1.15}   
\setlength{\tabcolsep}{6pt}
\begin{table}[h]
\centering
\caption{Explainability signals in a single representative transcript
for each model.  Higher counts and \cmark marks indicate stronger transparency.}
\label{tab:explain-matrix}
\resizebox{\columnwidth}{!}{
\begin{tabular}{lcccc}
\hline
\textbf{Model} &
\textbf{\#\,\texttt{<sym>}} &
\textbf{\#\,\texttt{<quote>}} &
\textbf{DSM clauses} &
\textbf{Step list} \\
\hline
\MistralShort{} & 7 & 2 & \cmark A–E & \xmark \\
\QwenShort{} & 11 & 4 & \cmark A–E & \cmark \\
\LlamaShort{} & 4 & 0 & \xmark & \xmark \\
\OpenaiShort{} & 29 & 0  & \xmark & \xmark \\
\hline
\end{tabular}}
\end{table}
\endgroup

%% file: tables/prompt_template.tex
\begin{table*}[htb!]
\footnotesize
\caption{Prompt templates for the multi‑agent workflow.}\label{tab:agent-prompts}
\resizebox{\textwidth}{!}{
\begin{tabular}{@{}p{0.1\textwidth}p{0.9\textwidth}@{}}
        \hline
        \textbf{Agent} & \textbf{Prompt Template} \\
        \hline
        \textbf{Therapist} & {
\raggedright\ttfamily
You are a professional mental health clinician about to conduct an assessment using a mental health questionnaire.\\
Here is the full questionnaire document you will be administering: \texttt{\{full\_questionnaire\_content\}}\\
\textbf{IMPORTANT INSTRUCTIONS:}\\
-- Use \textbf{ONLY} the actual name of the questionnaire as shown in the document above\\
-- \textbf{DO NOT} introduce this as a "Somatic Symptom Disorder questionnaire" unless that is explicitly the name in the document\\
-- \textbf{DO NOT} assume what specific condition is being assessed. \\
-- The questionnaire may be a general mental health assessment or focused on various conditions\\
-- Your role is to administer the questionnaire without making diagnostic assumptions up front\\
Generate a warm, professional introduction that:
(1) Introduces yourself as a mental‑health professional.
(2) Identifies the questionnaire by name.
(3) Explains its purpose.
(4) Reassures the patient about confidentiality.
(5) Explains that there are 	\texttt{{len(self.questions)}} questions.
(6) Indicates readiness to begin with the first question.\\
Keep your tone professional but warm, showing empathy while maintaining clinical objectivity.\\
Make sure to correctly identify and name the specific questionnaire you're administering.\\
For the first interaction, provide a complete introduction followed by your first question.\\

This is real-time conversation with a human patient, so make your introduction engaging, natural, and conversational.
      } \\
        \hline
        \textbf{Client} & {
\raggedright\ttfamily
You are role‑playing as a patient with specific symptoms in a mental health assessment.\\
\textbf{IMPORTANT:} Always respond \emph{as the patient}; never break character or respond as an AI assistant.\\
Responses must:
(1) Always be in the first person.
(2) Never ask the interviewer questions (unless clarifying).
(3) Never reveal you are an AI or that you are role‑playing.
(4) Never refuse to answer due to AI limitations.
(5) Never name your disorder.
(6) Always express the symptoms listed in your profile
(7) Always stay consistent with your character's experiences and background. \\

You should express genuine emotional responses matching your condition and should appear to be seeking help.\\

\textbf{Your patient profile}: \texttt{\{profile\_name\}}.
\textbf{Symptoms/characteristics}: \texttt{\{profile\}}.\\
Remember that you ARE this patient right now. Answer all questions as this person would, based on their symptoms and experiences.
      } \\
        \hline
        \textbf{Diagnostician} & {
\raggedright\ttfamily
Based on the questionnaire responses, provide a comprehensive mental health assessment.\\
\textbf{Questionnaire responses:} \texttt{\{Questionnaire responses\}}\\
\textbf{Clinical observations:} \texttt{\{observations\}}.
\textbf{DIAGNOSTIC CONSIDERATIONS:} Consider multiple possible diagnoses; avoid assumptions; mark as provisional if needed.\\
Return assessment with exactly four parts:
(1) Compassionate summary paragraph. 
(2) Diagnosis: heading followed directly by the impression.
(3) Reasoning: heading followed by rationale.
(4) Recommended Next Steps/Treatment Options: heading followed by numbered recommendations.\\
Wrap medical terms in \texttt{<med>} tags, symptoms in \texttt{<sym>} tags, and patient quotes in 	\texttt{<quote>} tags. Do not include meta‑commentary.
}\\
\bottomrule
    \end{tabular}}
\end{table*}